\documentclass[prb,showpacs,amsmath,amssymb,preprintnumbers,twocolumn]{revtex4}

\usepackage{graphicx}
\usepackage{dcolumn}
\usepackage{bm}

\preprint{submitted to Journal of Applied Physics}

\begin{document}

\title{Influence of mixing the low-valent transition metal atoms (Y,Y$^*$=Cr,Mn,Fe)  on the properties of the quaternary
 Co$_2$[Y$_{1-x}$Y$^*_x$]Z  (Z=Al,Ga,Si,Ge,Sn) Heusler compounds}

\author{K. \"Ozdo\~gan}\email{kozdogan@gyte.edu.tr}
\affiliation{Department of Physics, Gebze Institute of Technology,
Gebze, 41400, Kocaeli, Turkey}

\author{I. Galanakis}\email{galanakis@upatras.gr}
\affiliation{Department of Materials Science, School of Natural
Sciences, University of Patras,  GR-26504 Patra, Greece}

\author{E. \c Sa\c s\i o\~glu}\email{e.sasioglu@fz-juelich.de}
\affiliation{Institut f\"ur Festk\"orperforschung,
Forschungszentrum J\"ulich, D-52425 J\"ulich, Germanyy\\
Fatih University,  Physics Department, 34500,    B\" uy\" uk\c
cekmece,  \.{I}stanbul, Turkey}

\author{B. Akta\c s}
\affiliation{Department of Physics, Gebze Institute of Technology,
Gebze, 41400, Kocaeli, Turkey}

\date{\today}

\begin{abstract}
We complement our study on the doping and disorder in Co$_2$MnZ
compounds [I. Galanakis \textit{et al.}, Appl. Phys. Lett.
\textbf{89}, 042502 (2006) and K.  \"Ozdo\~gan \textit{et al.},
Phys. Rev. B \textbf{74}, (2006)] to cover also the quaterarny
Co$_2$[Y$_{1-x}$Y$^*_x$]Z compounds with the lower-valent
transition metals Y,Y$^*$ being Cr, Mn or Fe and the sp atom Z
being one of Al, Ga, Si, Ge, Sn. This study gives a global
overview of the magnetic and electronic properties of these
compounds since we vary both Y and Z elements. Our results suggest
that for realistic applications the most appropriate compounds are
the ones belonging to the families Co$_2$[Mn$_{1-x}$Cr$_x$]Z with
$x>0.5$ irrespectively of the nature of the $sp$ atoms since they
combine high values of majority DOS at the Fermi level due to the
presence of Cr, and half-metallicity with large band-gaps. On the
other hand the presence of Fe lowers considerably the majority
density of states at the Fermi level and when combined with an
element belonging to the Si-column, it even can destroy
half-metallicity.
\end{abstract}


\maketitle

\section{Introduction\label{sec1}}

Spintronics also known as magnetoelectronics is the newest growing
branch of magnetism.\cite{Zutic} The main idea is to replace
conventional electronics by a new kind of devices where the
central role is played by the spin of the electrons and not the
charge itself. This will allow the achievement of very low energy
consumption in combination with other desirable features like the
non-volatility for magnetic random-access memories (MRAMS), which
have recently found application in automotive industry. The
emergence and rapid growing of this research area brought in the
center of scientific research the so-called half-metallic
ferromagnets due to their possible
applications.\cite{book,Reviews,Westerholt,Reiss,Sakuraba,Dong}
These materials are hybrids between metals and semiconductors or
insulators, presenting metallic behavior for one spin-band and
semiconducting for the other, and thus overall they are either
ferro- or ferrimagnets with perfect spin-polarization at the Fermi
level. de Groot and his collaborators in 1983 were the first to
predict the existence of half-metallicity in the case of the
intermetallic Heusler alloy NiMnSb.\cite{deGroot} Several new
half-metallic ferromagnetic materials and their properties have
been initially predicted by theoretical ab-initio calculations and
later verified by experiments.

Although the Heusler alloys like NiMnSb (known as half- or
semi-Heusler compounds) have monopolized initially the interest,
the last approximately five years the interest has been shifted to
the so called full-Heusler compounds and mainly to the ones
containing Co, like Co$_2$MnAl. These alloys date from 1971, when
Webster managed to synthesize the first Heusler alloys containing
cobalt,\cite{Webster} and in early 90's it was argued in two
papers by a japanese group that they should be
half-metals.\cite{Ishida-Fujii} These primary calculations paved
the way and state-of-the-art first-principles calculations by
Picozzi et al\cite{Picozzi} and Galanakis et al\cite{Galanakis} in
2002 confirmed the predictions of Ref.~\onlinecite{Ishida-Fujii}.
Early experiments on these alloys were focused exclusively on the
growth of the full-Heusler alloys,\cite{growth} while latter
experimental studies have also dealt the magnetic properties of
the films,\cite{magnetism} the role of defects and
antisites,\cite{defects} superlattices,\cite{superlattices}
transport properties\cite{transport} and even the most complex
subject of their incorporation in realistic devices.\cite{devices}

First principles calculations have been extensively employed to
study the properties of the full-Heusler alloys. Several Heusler
alloys have been shown to be
half-metallic.\cite{Galanakis,Podlucky}  Ab-initio calculations
have shown that the surfaces\cite{Gala-Surf} and interfaces of the
full-Heusler compounds\cite{Interfaces} loose their
half-metallicity but Hashemifar and collaborators have shown that
it is possible to restore half-metallicity at some
interfaces.\cite{Hashemifar} Except interfaces states also
temperature driven excitations\cite{Chioncel,MavropTemp,Skomski}
and defects\cite{Picozzi04} seem to destroy half-metallicity. Also
some other important aspect of these alloys like the orbital
magnetism,\cite{orbit} the structural stability\cite{Block} and
the interplay of exchange interactions\cite{Sasioglu,Kurtulus}
have been addressed in literature.

Over the last two years the interest in full-Heusler alloys
containing cobalt has been focused on the so-called quaternary
Heusler alloys, which are found to present half-metallicity as
long as the corresponding perfect compounds are
half-metals.\cite{GalaQuat} The material of choice was
Co$_2$[Cr$_{1-x}$Fe$_{x}$]Al. The half-metallicity has been
predicted for all concentrations $x$,\cite{Antonov} and Miura et
al have studied theoretically the stability of these compounds
versus the creation of defects and antisites.\cite{Miura} Also the
family Co$_2$[Mn$_{1-x}$Fe$_{x}$]Si has been extensively
studied\cite{Balke} due to the fact that ab-initio calculations,
including the on-site Coulomb repulsion (the so-called Hubbard
$U$), have shown that Co$_2$FeSi can reach a total spin magnetic
moment of 6 $\mu_B$ which is the largest known spin moment for a
half-metal.\cite{Kandpal,Wurmehl} Several experiments have been
devoted to the study of the structural and magnetic properties of
these quaternary Heusler alloys\cite{Elmers} and such films have
been incorporated both in magnetic tunnel junctions\cite{Marukame}
and spin-valves.\cite{Kelekar}

\begin{figure}
\includegraphics[scale=0.45]{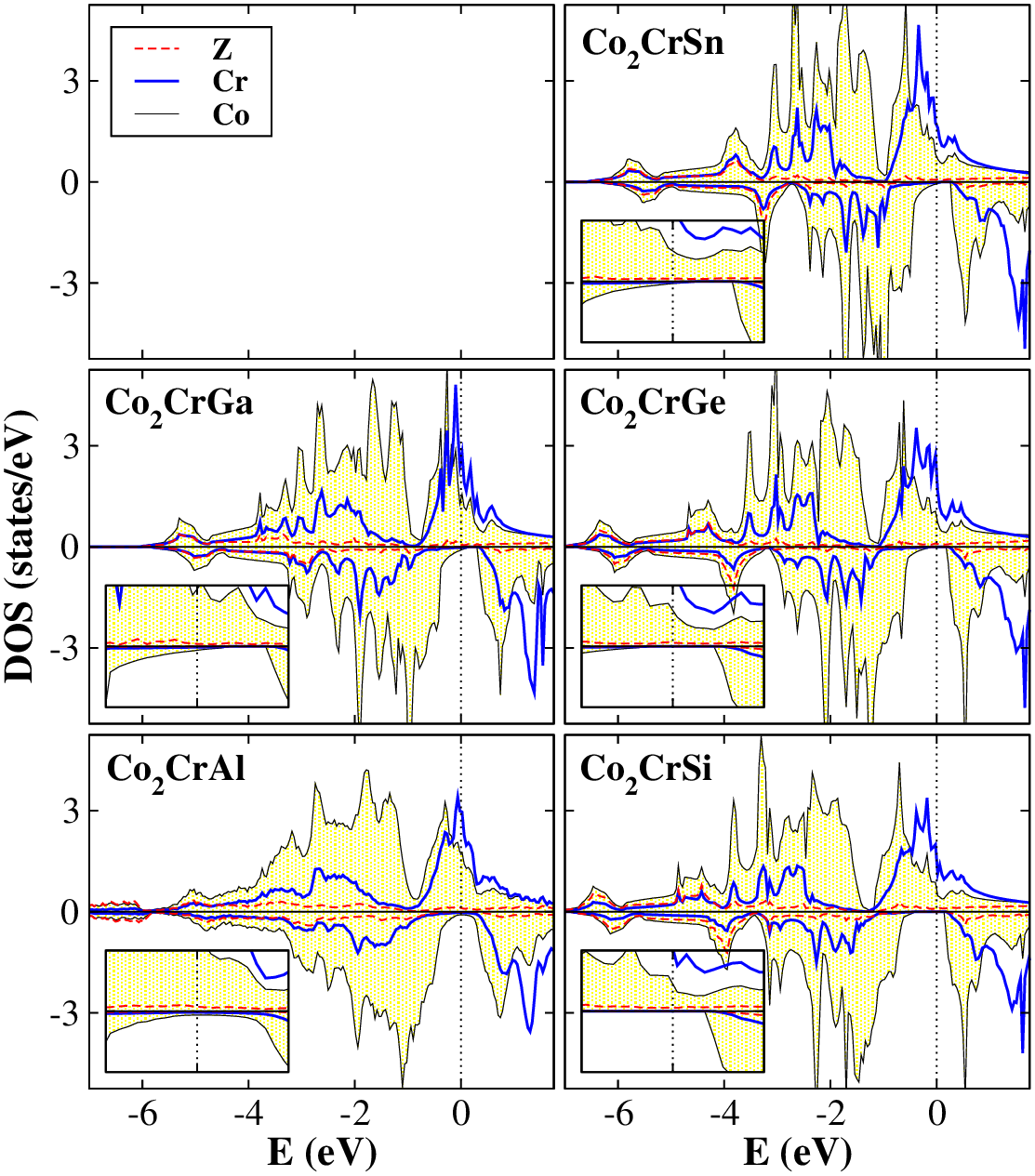}
\caption{(Color online) Atom-resolved density of states (DOS) for
the Co$_2$CrZ compounds, where Z is Al, Ga, Si, Ge, and Sn.  We
have set the Fermi level as the zero of the Energy axis. In the
onsets we have blown up the region around the Fermi level. Note
that positive values of DOS refer to the majority-spin electrons
and negative values to the minority-spin electrons. \label{fig1}}
\end{figure}

\section{Description of present calculations}

In a recent paper,\cite{APL} we employed the full--potential
nonorthogonal local--orbital minimum--basis band structure scheme
(FPLO)\cite{koepernik} to study the effect of doping and disorder
on the magnetic properties of the Co$_2$MnSi, Co$_2$MnGe,
Co$_2$MnSn full-Heusler alloys. Doping simulated by the
substitution of Cr and Fe for Mn in these alloys overall kept the
half-metallicity. The effect of doping depended clearly on the
position of the Fermi level, having the largest one in the case of
Co$_2$MnSi where the Fermi level is near the edge of the
minority-spin gap. Latter we expanded this work to cover also the
case of Co$_2$MnAl and Co$_2$MnGa compounds\cite{PRB-BR} which
have one valence electron less than the previous ones. Also in
that case a high degree of spin-polarization at the Fermi level
was overall preserved. Finally for all five compounds we found
that the creation of antisites severely affects the half-metallic
character of the compounds.

\begin{figure}
\includegraphics[scale=0.45]{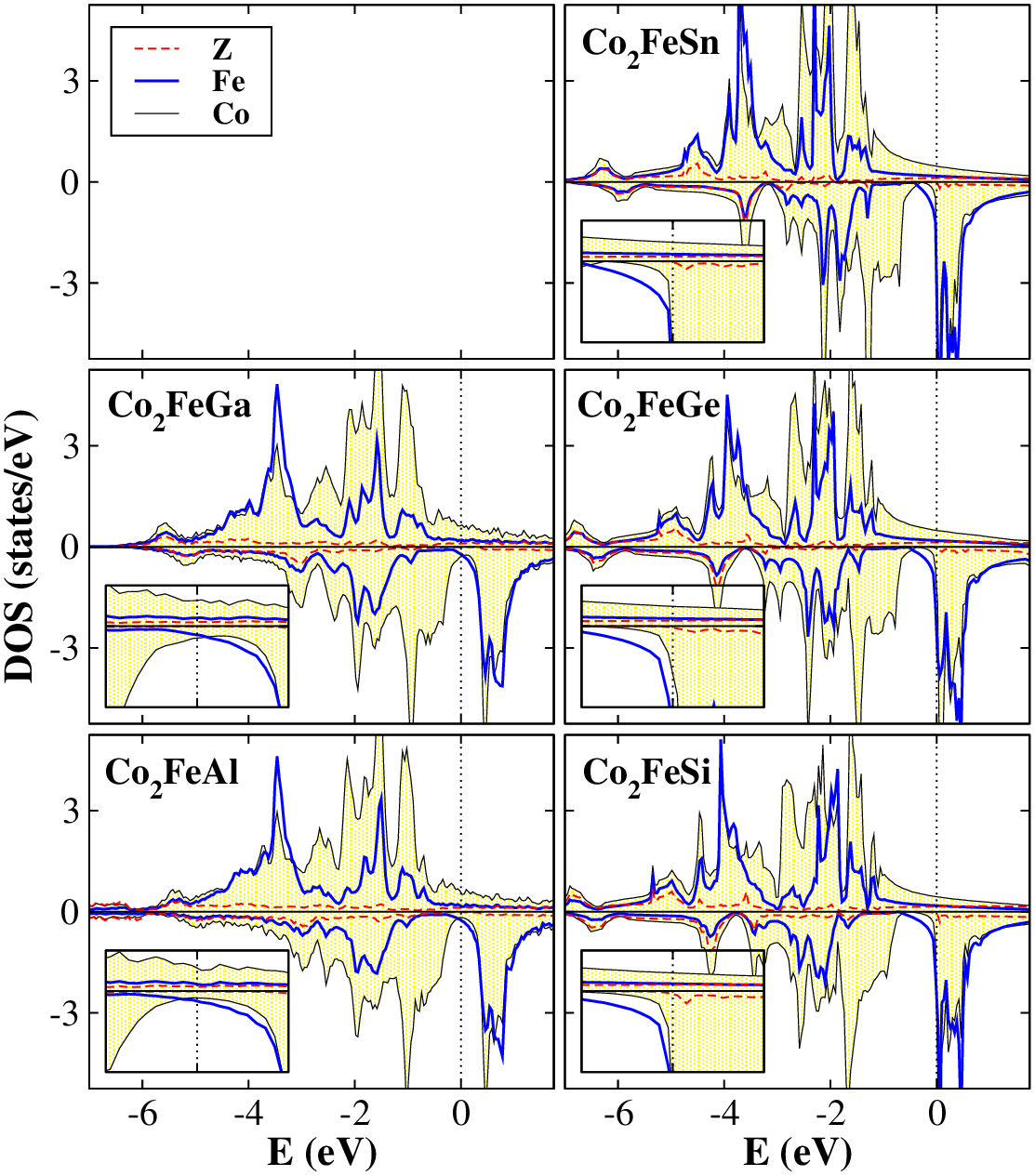}
\caption{(Color online) Same as Fig. \ref{fig1} for the Co$_2$FeZ
compounds where Z is Al, Ga, Si, Ge, and Sn.
 \label{fig2}}
\end{figure}

In this manuscript we expand these two studies to cover not only
the case of doping but all the families of resulting quaternary
Heusler alloys; Co$_2$[Mn$_{1-x}$Cr$_{x}$]Z and
Co$_2$[Mn$_{1-x}$Fe$_{x}$]Z, with Z being Al, Ga, Si, Ge, Sn. For
reasons of completeness we have decided to calculate also the case
when we mix Cr and Fe atoms at the site occupied by the
lower-valent transition metal atoms; Co$_2$[Cr$_{1-x}$Fe$_{x}$]Z
families.  Since we change both the transition-metal atoms  and
the sp atoms, we get a global feeling of the behavior of the
magnetic and electronic properties of these compounds. We have
employed the FPLO electronic structure method as already stated in
conjunction with the local-spin-density approximation (LSDA). The
coherent potential approximation (CPA) was used to simulate the
disorder. We have used the experimental lattice constants for the
perfect compounds containing Mn\cite{lattice,lattice2} and we have
kept them constant when substituting with Fe or Cr since no
evidence is known on the exact behavior of the quaternary
compounds. This tactic is different than the one used in
Ref.~\onlinecite{GalaQuat} where it was assumed that the lattice
constant varies linearly with the concentration. For the cases
under study both methods give lattice constants within less than
1\% difference and thus practically identical results. Finally we
should note that we discuss half-metallicity in terms of total
spin-moments since perfect half-metals show the Slater-Pauling
(SP) behavior; the total spin moment in the unit cell is the
average number of valence electrons minus 24.\cite{Galanakis}
Spin-polarization at the Fermi level is not considered since it
remains very close to the perfect 100\%  and its exact value
depends on computational details
 contrary to the total spin moments which were found to be more robust.

Before presenting our results we have drawn in Figs. \ref{fig1}
and \ref{fig2} the atom-resolved DOS for the Co$_2$CrZ and
Co$_2$FeZ compounds. On the left column of each figure are the
cases with Z a $sp$ element belonging to the IIIB column of the
periodic table (Al and Ga) and in the right column the cases of
IVB elements (Si, Ge and Sn). For the Co$_2$CrZ alloys the extra
electron in the latter case occupies majority states leading to an
increase of the exchange splitting between the occupied majority
and the unoccupied minority states and thus to larger gap-width
for the Si-, Ge- and Sn-based compounds with respect to the Al-and
Ga-based alloys; a similar behavior was present also for the
Co$_2$MnZ compounds.\cite{PRB-BR} In the case of the compounds
containing Fe the extra electrons with respect to Cr and Mn
compounds lead to an overlap  of the Co bonding and antibonding
minority d-hybrids and the gap is already destroyed for the Co
atoms. Moreover the unoccupied Fe states move lower in energy
since the majority occupied Fe states are very deep in energy. The
Al and Ga compounds keep a high degree of spin-polarization but
the phenomenon is very intense for the Si, Ge and Sn compounds,
which have 30 electrons per unit cell, and half-metallicity is
destroyed. As argued by Kandpal et al, in the latter case
(contrary to the Mn and Cr cases) the on-site correlations play a
drastic role and taking them into account would push the Fe
minority states higher in energy leading again to the opening of
the gap in the minority density of states (DOS).\cite{Kandpal} The
same effect occurs also if we expand the lattice constant by ~5
\%, where the Fermi level is pushed deeper in the
energy.\cite{Podlucky} Unfortunately the scheme used in
Ref.~\onlinecite{Kandpal} cannot be easily employed when disorder
is present and it implicates the use of an ad-hoc parameter $U$
(the so-called Hubbard parameter) which is not easily determined,
and thus we do not take into account the on-site correlation
effects in our study.

\section{Substituting C\lowercase{r} or F\lowercase{e} for M\lowercase{n} in C\lowercase{o}$_2$M\lowercase{n}Z
alloys\label{sec2}}

\begin{figure}
\includegraphics[scale=0.45]{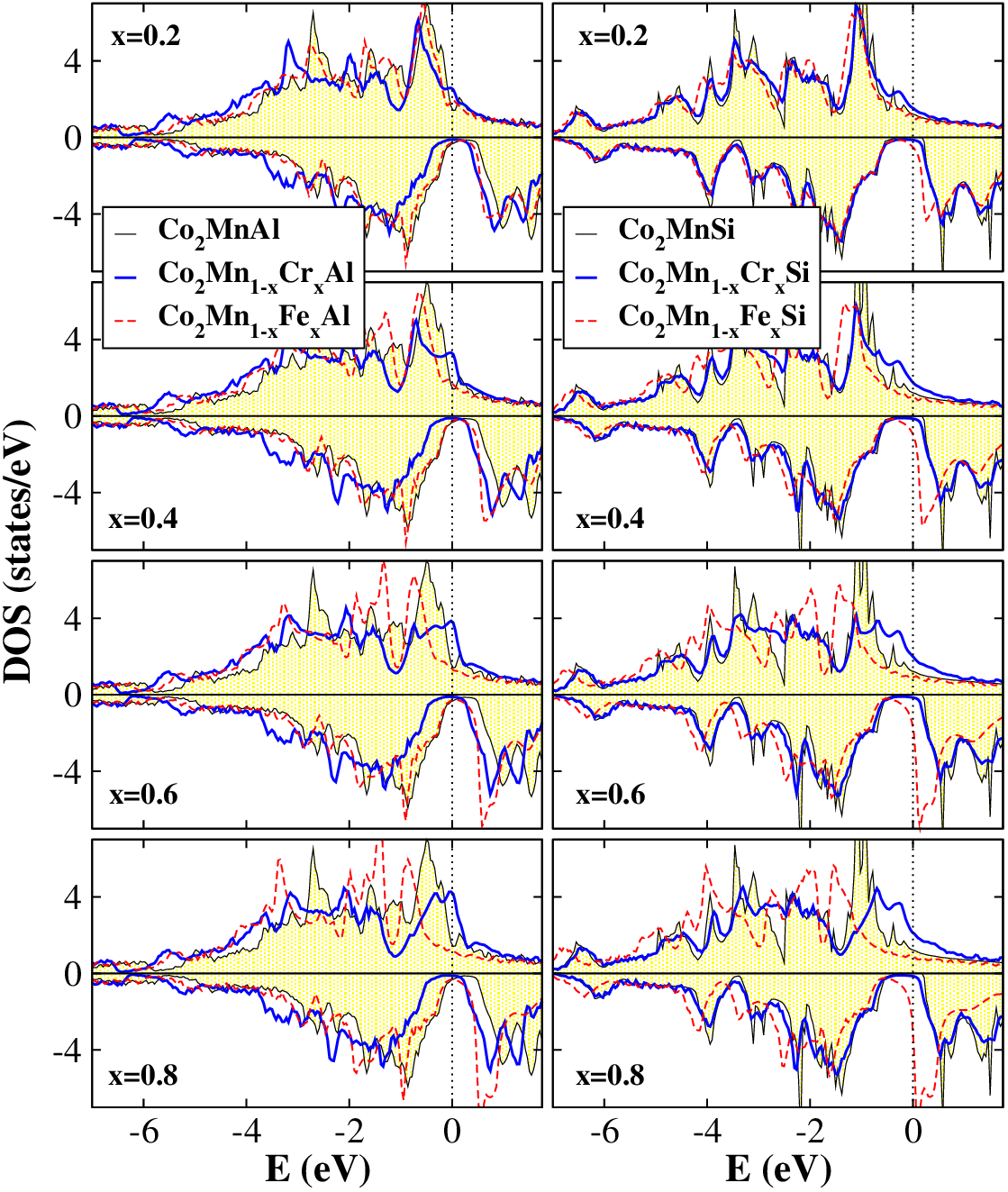}
\caption{(Color online) Total DOS in the case of the
Co$_2$[Mn$_{1-x}$(Cr or Fe)$_x$]Al  (left panel) and
Co$_2$[Mn$_{1-x}$(Cr or Fe)$_x$]Si compounds  (right panel) for
four different values of the concentration $x$. The total DOS are
compared with the ones of the perfect Co$_2$MnAl and Co$_2$MnSi
compounds. Details as in Fig. \ref{fig1}. \label{fig3}}
\end{figure}

We will start our discussion from the two families of compounds
containing both Co and Mn, where we substitute Fe and Cr for Mn.
Substitution of Fe for Mn represents an increase of the total
number of valence electrons, while the vice versa is true for the
substitution of Cr for Mn. In Fig.~\ref{fig3} we have drawn the
total DOS for the Co$_2$[Mn$_{1-x}$(Cr or Fe)$_x$]Al on the left
column and for the Si-instead-of-Al compounds (right column) with
respect to the perfect Co$_2$MnAl or Co$_2$MnSi alloys for
different values of the concentration $x$.

As we stated in the previous section the Co$_2$CrAl and Co$_2$CrSi
compounds are half-metals showing similar behavior to the
Co$_2$MnAl and Co$_2$MnSi compounds. Thus also the intermediate
quaternary compounds are half-metals. If the Z atom is Al or Ga,
as we replace Cr for Mn, there is a leakage of charge from the
majority states near the Fermi level towards the unoccupied
majority states to account for the decrease in the average number
of valence electrons. In the same time the change in the majority
states influences also the exchange splitting and the minority
bands move as in a rigid band lower in energy but the Fermi level
remains within the gap (the minority blue line representing the
quaternary compound moves lower in energy with the concentration
with respect to the shaded region surrounded by the black line
representing the ideal Co$_2$MnAl alloy). In the case of the Si,
Ge and Sn (in the figure we show only the Si case) compounds, the
alloys have one more valence electron and thus the exchange
splitting between majority occupied Cr-Mn d-states and minority
unoccupied Cr-Mn d-states is stronger and as we substitute Cr for
Mn the minority states are almost identical to the one of the
perfect Co$_2$MnSi alloy and only the majority states just below
the Fermi level show a leakage towards unoccupied states to
account for the decrease in the electron charge.

\begin{figure}
\includegraphics[scale=0.45]{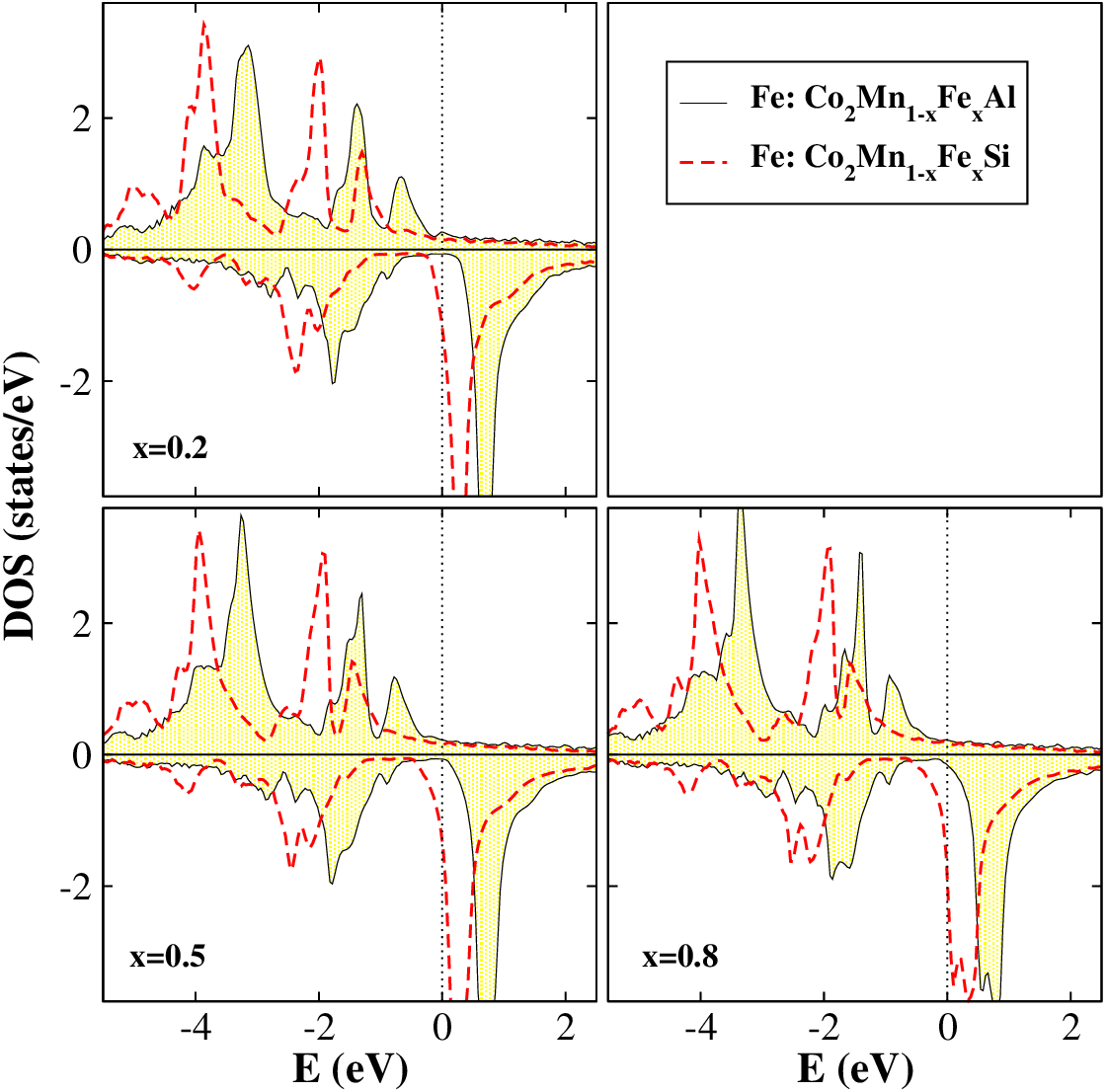}
\caption{(Color online) Fe-resolved DOS in the case of the
Co$_2$[Mn$_{1-x}$Fe$_x$]Al and Co$_2$[Mn$_{1-x}$Fe$_x$]Si
compounds for three different values of the concentration $x$.
Details as in Fig. \ref{fig1}. \label{fig4}}
\end{figure}

The reverse phenomena are present for the substitution of Fe for
Mn. In the case of the Co$_2$[Mn$_{1-x}$F$_x$]Al alloys the number
of valence electrons is larger than for the
Co$_2$[Mn$_{1-x}$Cr$_x$]Al ones and they show behavior similar to
the Co$_2$[Mn$_{1-x}$Cr$_x$]Si family, showing small variations
for the minority occupied states and the extra charge due to the
increase in the Fe-concentration pushes some majority states below
the Fermi level. The Co$_2$[Mn$_{1-x}$Fe$_x$]Al compounds are
half-metal for all the values of the concentration $x$ although
the gap is smaller than in the corresponding alloy with Cr-Mn
mixing. The case of the compounds containing the heavier sp atoms
like Si is more difficult since now already for the perfect
Co$_2$MnSi there are 29 valence electrons. For the Co$_2$FeSi we
have to fill 18 majority states (there are 12 minority occupied
ones for the half-metallic full-Heusler alloys). This is difficult
energetically since we have to occupy the high lying antibonding
majority states and the system prefers to  loose half-metallicity
to gain energy. Kandpal and collaborators have argued that in this
case  the on-site correlation effects, which are not taken into
account by conventional LSDA calculations, become important and
they can reopen the gap and restore the
half-metallicity.\cite{Kandpal} To make this discussion more clear
we have drawn in Fig. \ref{fig4} the Fe-resolved DOS for both
Co$_2$[Mn$_{1-x}$Fe$_x$]Al and Co$_2$[Mn$_{1-x}$Fe$_x$]Si
compounds for three different values of the concentration $x$. All
graphs have been scaled to one atom. For the Al-based alloy Fe
shows identical DOS irrespectively of its concentration in the
quaternary alloy. On the other hand in the Si-based compound as
the concentration of Fe in the alloy increases the extra charge
occupies also minority states above the gap which now cross the
Fermi level.

\begin{figure}
\includegraphics[scale=0.45]{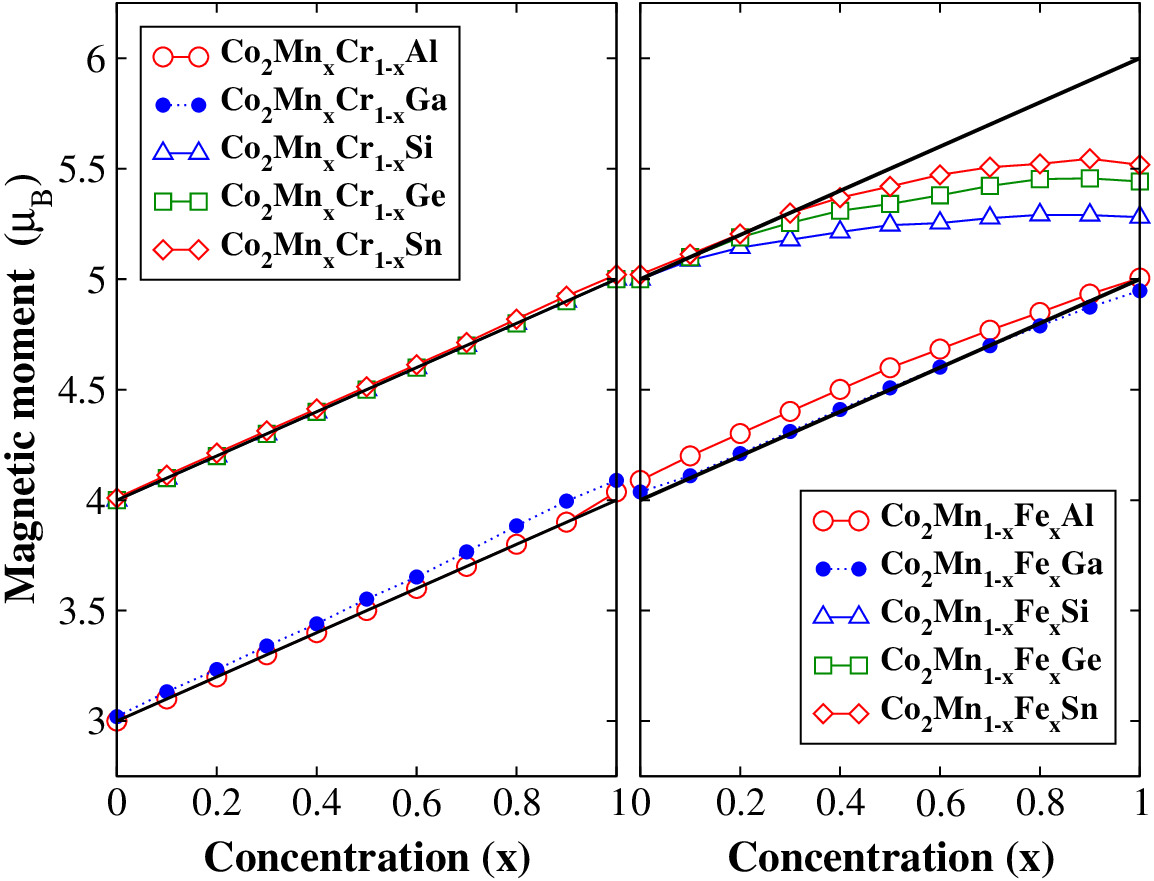}
\caption{(Color online) Total spin magnetic moment as a function
of the concentration $x$ for the studied Co$_2$[Mn$_{1-x}$Cr$_x$]Z
and Co$_2$[Mn$_{1-x}$Fe$_x$]Z Heusler compounds.The solid black
lines represent the Slater-Pauling behavior. \label{fig5}}
\end{figure}

The above discussion on the total DOS and the question of the
preservation of the half-metallicity in the quaternary compounds
is reflected on the total spin moments and in Fig.~\ref{fig5} we
have drawn the variation of the total spin moment as a function of
the concentration $x$ for all compounds under study in this
section. The solid black lines represent the Slater-Pauling
behavior obeyed by the perfect half-metallic ferromagnets. It is
obvious that when the sp atom is either Al or its isoelectronic Ga
the compounds show behavior very close to the SP one and thus
although not perfect half-metals they present very high values of
spin-polarization at the Fermi level. On the other hand compounds
containing the heavier Si, Ge and Sn atoms are perfect half-metals
when we mix Cr and Mn, but when we substitute Fe for Mn the
half-metallicity is lost even for a concentration of Fe of 0.2.
For concentration values between 0.4 and 1 the
Co$_2$[Mn$_{1-x}$Fe$_x$](Si, Ge or Sn) total spin moments show a
plateau being constant. We will discuss the smaller deviations for
the Sn and Ge compounds with respect to the Si one in the next
section since the same occurs also for the
Co$_2$[Cr$_{1-x}$Fe$_x$](Si, Ge or Sn) alloys.  Now the question
arises what is the mechanism for this. In Table~\ref{table1} we
have gathered the atom-resolved spin moments for all compounds
under study including also the Co$_2$[Cr$_{1-x}$Fe$_x$]Z families.
As we can see the Fe-resolved spin moment in the case of the
Co$_2$[Mn$_{1-x}$Fe$_x$]Si compounds is almost constant because
the extra charge occupies equally majority and minority states and
the number of uncompensated spin is almost constant. Since no
increase in Fe spin-moment occurs and the Fe spin-moment is
considerably smaller than the Mn one the total spin moment is not
increased.

\begin{table}
\caption{Total and atom-resolved spin magnetic moments for the
case of the studied Al and Si compounds  in $\mu_B$. The total
moment in the cell is the sum of the atomic ones multiplied by the
concentration of this chemical element. Note that for Cr, Mn and
Fe we have scaled the spin moments to one atom and that for Co we
give the sum of the moments of both atoms.} \label{table1}
\begin{ruledtabular}
 \begin{tabular}{l|ccccc|ccccc}
 $x$& \multicolumn{5}{c|}{Co$_2$[Cr$_{1-x}$Fe$_{x}$]Al} & \multicolumn{5}{c}{Co$_2$[Cr$_{1-x}$Fe$_{x}$]Si}  \\
   &    Total  &   Co &   Cr&Fe&
 sp &  Total & Co &   Cr&Fe&  sp \\ \hline
  0.00  &  3.00&   1.46& 1.63&      &  -0.09&4.00&   1.89&    2.17  &     &
  -0.09\\
  0.20  &  3.40&   1.72& 1.49& 2.89 &  -0.09&4.31&   2.04&    2.21 &    2.78    &
  -0.08\\
  0.40  &  3.80&   1.87& 1.47& 2.86 &  -0.10&4.57&   2.18&    2.24&    2.75    &
  -0.08\\
  0.60  &  4.19&   2.02& 1.48& 2.83 &  -0.11&4.82&   2.31&    2.25 &    2.75    &
  -0.07\\
  0.80  &  4.58&   2.15& 1.50& 2.80 &  -0.12&5.05&   2.43&    2.28  &    2.75    &
  -0.07\\
  1.00  &  4.95&   2.27&     & 2.79 &  -0.12&5.28&   2.52&           &  2.79       &
  -0.06\\ \hline
$x$& \multicolumn{5}{c|}{Co$_2$[Mn$_{1-x}$Cr$_{x}$]Al} & \multicolumn{5}{c}{Co$_2$[Mn$_{1-x}$Cr$_{x}$]Si}  \\
   &    Total  &   Co &   Mn&Cr&
 sp &  Total & Co &   Mn&Cr&  sp \\ \hline
  0.00  &  4.04& 1.36& 2.82&        &  -0.14&5.00&   1.96&    3.13  &     &
  -0.09\\
  0.20  &  3.80& 1.54& 2.74& 0.91   &  -0.11&4.80&   1.97&     3.12 &    2.09    &
  -0.08\\
  0.40  &  3.60&   1.55& 2.77& 1.20 &  -0.10&4.60&   1.95&     3.12&    2.12    &
  -0.08\\
  0.60  &  3.40&   1.54& 2.79& 1.37 &  -0.09&4.40&   1.93&     3.13 &    2.15    &
  -0.07\\
  0.80  &  3.20&   1.53& 2.83& 1.48 &  -0.08&4.20&   1.91&     3.13  &    2.17    &
  -0.07\\
  1.00  &  3.00&   1.46&     & 1.63 &  -0.09&4.00&   1.89&    &  2.17       &
  -0.06\\ \hline
$x$& \multicolumn{5}{c|}{Co$_2$[Mn$_{1-x}$Fe$_{x}$]Al} & \multicolumn{5}{c}{Co$_2$[Mn$_{1-x}$Fe$_{x}$]Si}  \\
   &    Total  &   Co &   Mn&Fe&
 sp &  Total & Co &   Mn&Fe &  sp \\ \hline
  0.00  &  4.04&   1.36& 2.82&      &  -0.14&5.00&   1.96&    3.13  &     &
  -0.09\\
  0.20  &  4.21&   1.58& 2.76& 2.79 &  -0.13&5.14&   2.13&     3.15 &    2.82    &
  -0.08\\
  0.40  &  4.41&   1.76& 2.78& 2.79 &  -0.13&5.21&   2.25&     3.18&    2.79    &
  -0.07\\
  0.60  &  4.60&   1.94& 2.82& 2.79 &  -0.13&5.25&   2.36&     3.20 &    2.78    &
  -0.05\\
  0.80  &  4.79&   2.12& 2.86& 2.79 &  -0.13&5.29&   2.46&     3.23  &    2.78    &
  -0.04\\
  1.00  &  4.95&   2.27&     & 2.79 &  -0.12&5.28&   2.52&            &  2.79      &
  -0.03
\end{tabular}
\end{ruledtabular}
\end{table}

\begin{figure}
\includegraphics[scale=0.45]{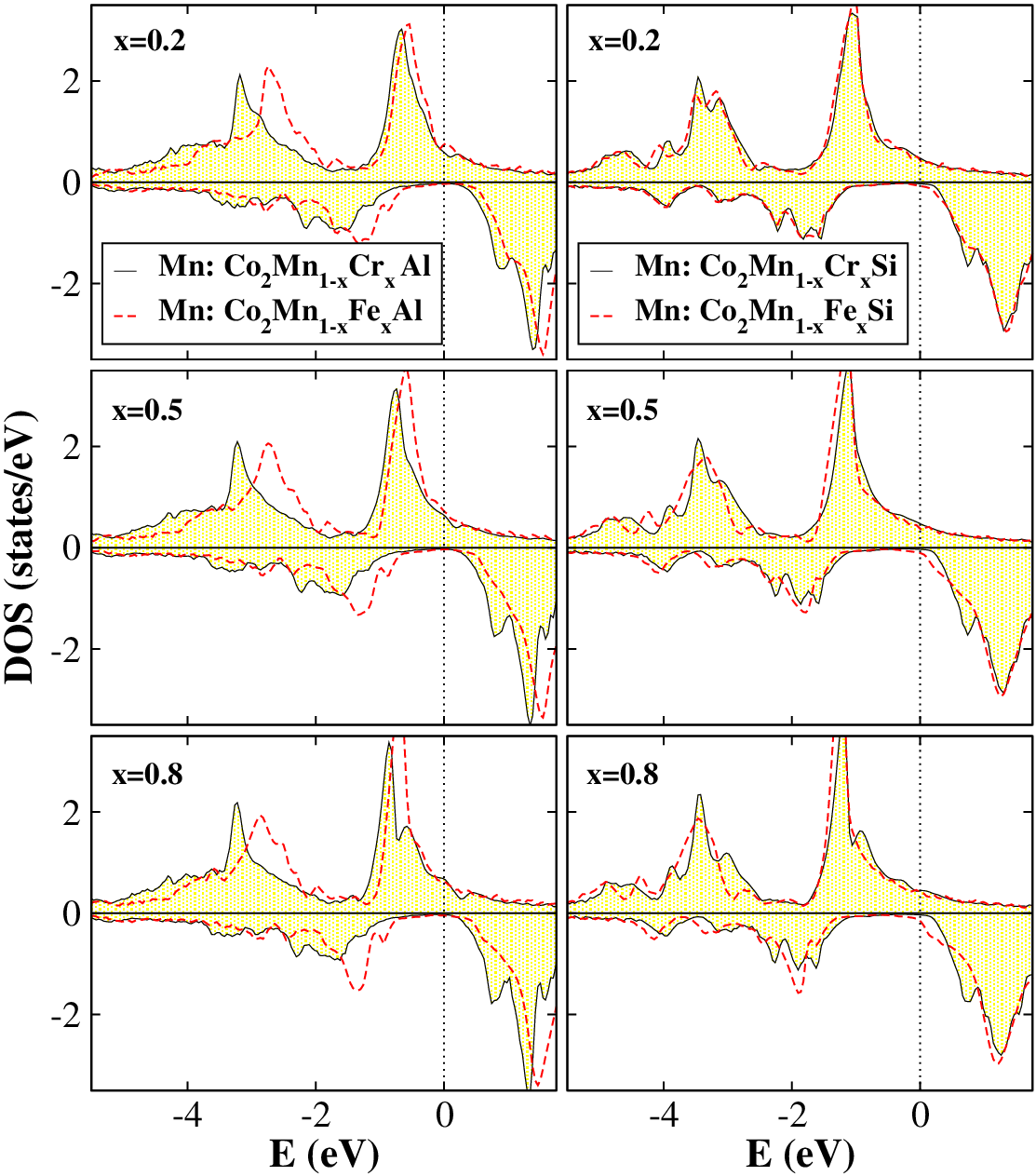}
\caption{(Color online) Mn-resolved DOS in the case of the
Co$_2$[Mn$_{1-x}$(Cr or Fe)$_x$]Al (left panel) and
Co$_2$[Mn$_{1-x}$(Cr or Fe)$_x$]Si (right panel) compounds for
three different values of the concentration $x$. Details as in
Fig. \ref{fig1}. \label{fig6}}
\end{figure}

To complete this section we will briefly also discuss the
atom-resolved spin moments of the different constituents (we will
omit for the present any reference to the
Co$_2$[Cr$_{1-x}$Fe$_{x}$]Z alloys). Overall Mn, Fe and Cr atoms
which play the role of the lower-valent transition-metal atom show
stable constant spin moments irrespectively of the concentration.
Cr in the case of Al and Ga based compounds has smaller moments
than for the heavier Si, Ge or Sn-based compounds and this
behavior is largely due to the relative position of the Fermi
level with respect to the majority pick which it crosses. This
behavior has been extensively discussed in
Ref.~\onlinecite{PRB-BR}. Mn has larger spin moment by about ~0.3
$\mu_B$ when the sp atom belong to the Si-column instead of the
Al-one accounting for a small fraction of the extra electron. The
fact that Mn shows a pretty standard behavior is also seen in Fig.
\ref{fig6} where we have drawn the Mn-resolved DOS for several
cases. It is obvious that Mn DOS is similar for all cases under
study with slight variations mainly due to the different
environment created by the Co atoms (each Mn atom has eight Co
atoms as first neighbors) which carry also the information about
the sp atoms and the Cr or Fe ones at the other sites. There are
cases where the Mn-DOS is more spiky but this is entirely due to
small shifts of the localized-in-energy $e_g$ states with respect
to the more delocalized $t_{2g}$ ones.

\begin{figure}
\includegraphics[scale=0.45]{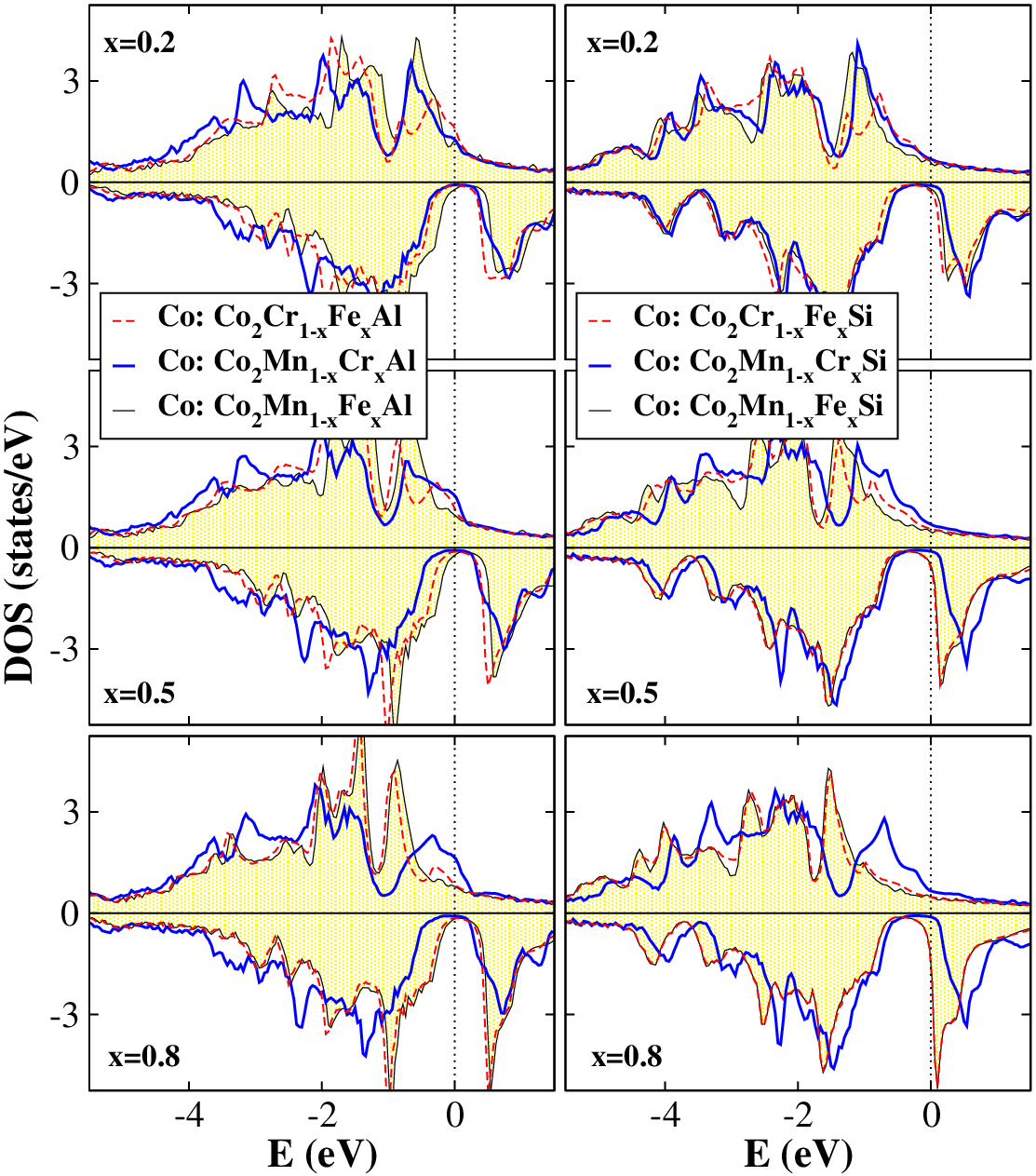}
\caption{(Color online) Co-resolved DOS in the case of the
Al-based (left panel) and Si-based (right panel) compounds.
Details as in Fig. \ref{fig1}. \label{fig7}}
\end{figure}

Co atoms show a more interesting behavior. As we substitute Fe for
Mn, Co atoms increase considerably their spin magnetic  moment to
account for the extra charge (the same occurs when we substitute
Fe for Cr), while when we dope with Cr the spin-moment of the Co
atoms is almost constant. Overall, as expected for the same type
of compounds the Co has a larger moment when the sp atom is a
Si-column one than a Al-column one due to the larger number of
valence electrons. In Fig.~\ref{fig7} we have drawn the
Co-projected DOS for all cases under study and for several values
of the concentration $x$. Every Co atom is surrounded by four
low-valent transition metal atoms (Cr, Mn or Fe) and four sp-atoms
carrying the information for the different systems through the
hybridization of the d-states of Co with the d-states of the other
transition-metal atoms or the p-states of the sp-atom. As it was
shown in Ref.~\onlinecite{Galanakis} the minority states just
around the gap (when there is a real gap) are exclusively located
at Co sites. Thus when we mix Mn with Cr, where a real gap exists
(DOS's with blue lines in the figure), the gap is fully determined
by the Co minority states. For high concentration of Cr in
Co$_2$[Mn$_{1-x}$Cr$_x$]Z, the Co atoms show a majority pick just
below the Fermi level due to the large polarization induced by the
Cr majority states at the Fermi level.\cite{Podlucky} In the case
of compounds where Fe is present, both Co$_2$[Mn$_{1-x}$Fe$_x$]Z
and Co$_2$[Cr$_{1-x}$Fe$_x$]Z, Fe minority states are near the
Fermi level (they even cross it) polarizing the Co d-states and
the gap of the Co minority states becomes more narrow. This
phenomenon is so intense, especially for large Fe concentrations,
that the Co-DOS close to the Fermi level is identical for both
families of compounds (red line for Co$_2$[Cr$_{1-x}$Fe$_x$]Z and
black shaded for Co$_2$[Mn$_{1-x}$Fe$_x$]Z). In general the
variation of the spin moment of the Co atoms can be used to
characterize the variation of the concentration in different
perfectly ordered samples.

\section{The case of the C\lowercase{o}$_2$C\lowercase{r}$_{1-x}$F\lowercase{e}$_x$Z family\label{sec3}}

\begin{figure}
\includegraphics[scale=0.45]{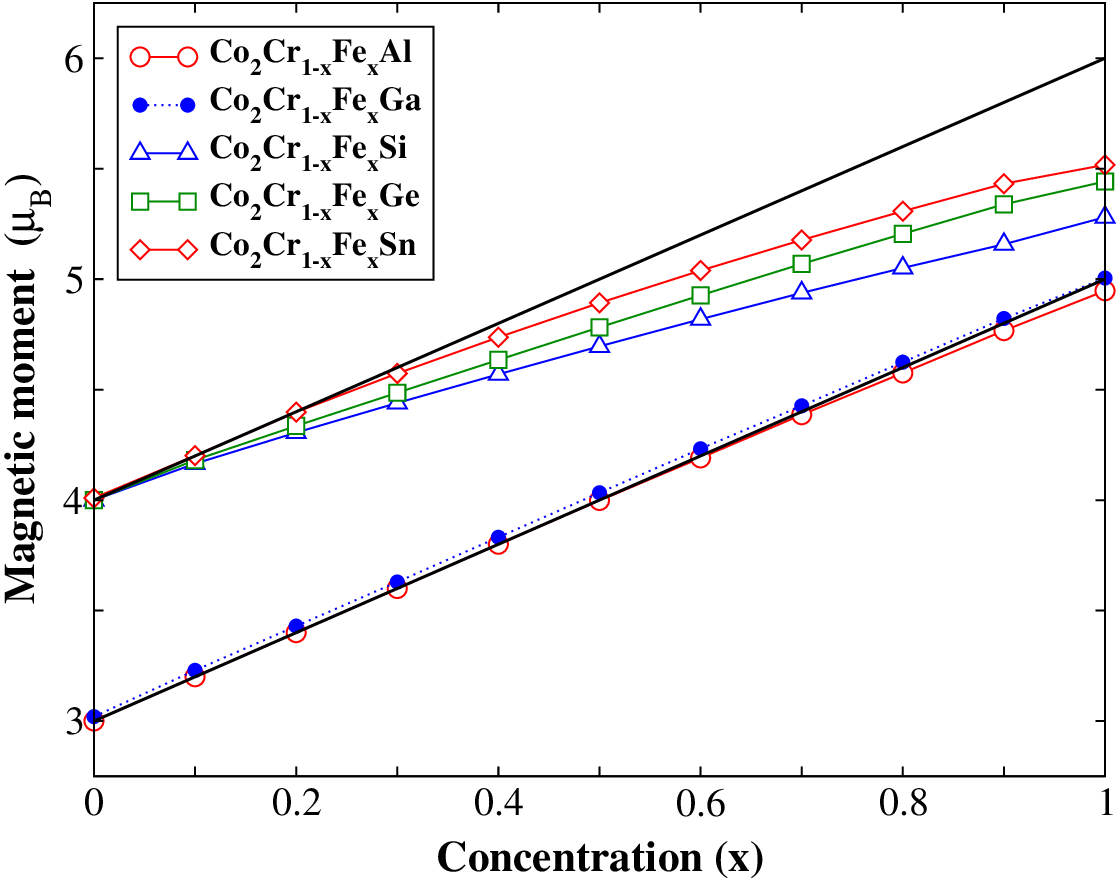}
\caption{(Color online) Total spin magnetic moment as a function
of the concentration $x$ for the studied Co$_2$[Cr$_{1-x}$Fe$_x$]Z
Heusler compounds.The solid black lines represent the
Slater-Pauling behavior. \label{fig8}}
\end{figure}

In the last part of our study we investigated the properties of
the Co$_2$[Cr$_{1-x}$Fe$_x$]Z alloys with Z being Al, Ga, Si, Ge
or Sn. As we have already mentioned above these compounds present
properties very similar to the Co$_2$[Mn$_{1-x}$Fe$_x$]Z families.
Fe plays a central role even at relatively low concentrations
affecting the half-metallicity. In Fig.~\ref{fig8} we have drawn
the behavior of the total spin moments with the concentration $x$.
The solid black lines represent the perfect SP behavior of the
ideal half-metallic ferromagnets. For the case where Z is either
Al or Ga the total spin moment should vary between 3 $\mu_B$ for
the alloy containing only Cr ($x=0$) and 5 $\mu_B$ for the alloy
containing exclusively Fe ($x=1$). The Ga-based compound falls
exactly on the solid black line and thus is half metallic for all
concentrations. The Al compounds deviates slightly only for very
high concentrations of Fe ($x$= 0.9 or 1). This is also observed
in Table~\ref{table1} where we present the atom-resolved and total
spin moments. For Co$_2$[Cr$_{0.2}$Fe$_{0.8}$]Al the total spin
moment is 4.58 instead of the ideal 4.6 and for Co$_2$FeAl the
total moment is 4.95 instead of the perfect value of 5. Of course
a slight increase of the lattice constant will push the Fermi
level deeper in energy and back in the gap and half-metallicity
will be restored even in these cases.

On the other the total spin moments, in the case where Z is Si or
its isoelectronic Ge and Sn, strongly deviate from the SP behavior
showing similar behavior to the case of
Co$_2$[Mn$_{1-x}$Fe$_{x}$](Si, Ge or Sn) compounds in
Fig.~\ref{fig5}. It is interesting that the Si compounds present
in general lower values of total spin moment than the Ge and Sn
ones in both cases (see Figs. \ref{fig5} and \ref{fig8}). To
elucidate this behavior we have included in Table \ref{table2} the
total and atom resolved spin moment for Co$_2$[(Cr or
Mn)$_{1-x}$Fe$_{x}$]-Si and -Sn compounds. The Co$_2$FeSi has a
total spin moment of 5.28 $\mu_B$ while Co$_2$FeSn shows a total
spin moment of 5.52 $\mu_B$. If we compare the trends between the
Si and Sn compounds, we remark that the Mn and Fe spin moments are
larger in the case of Z= Sn. Sn is a much heavier element than Si
and the valence $p$ electrons are  more extended in space but this
is overcompensated by the larger lattice constant. This larger
lattice constant leads also to much smaller spin moments at the Sn
site since a larger number of majority states is occupied than for
Si. Since the lattice constant is larger, Mn and Fe atoms have
more space around them and thus a larger Wigner-Seitz sphere and
their behavior becomes more atomic-like leading to an increase of
their atomic spin moments and thus to larger total spin moments.
The same phenomenon is also true for the compounds containing Ge.

\begin{table}
\caption{Total and atom-resolved spin magnetic moments for the
case of the studied Co$_2$[Cr$_{1-x}$Fe$_x$](Si or Sn)  in
$\mu_B$.} \label{table2}
\begin{ruledtabular}
 \begin{tabular}{l|ccccc|ccccc}
 $x$& \multicolumn{5}{c|}{Co$_2$[Cr$_{1-x}$Fe$_{x}$]Si} & \multicolumn{5}{c}{Co$_2$[Cr$_{1-x}$Fe$_{x}$]Sn}  \\
   &    Total  &   Co &   Cr&Fe&
 sp &  Total & Co &   Cr&Fe&  sp \\ \hline
  0.00  & 4.00&   1.89&    2.17  &           &  -0.09& 4.00&1.66&2.41& & -0.06   \\
  0.20  & 4.31&   2.04&    2.21 &    2.78    &  -0.08& 4.40&1.89&2.49&2.89& -0.06   \\
  0.40  &  4.57&   2.18&    2.24&    2.75    &  -0.08& 4.74&2.10&2.54&2.89&-0.05     \\
  0.60  &  4.82&   2.31&    2.25 &    2.75   &  -0.07& 5.04&2.30&2.60&2.90& -0.04   \\
  0.80  &  5.05&   2.43&    2.28  &    2.75  &  -0.07& 5.31&2.47&2.65&2.92& -0.02    \\
  1.00  &  5.28&   2.52&           &  2.79   &  -0.06& 5.52&2.61&
  &2.92& -0.01\\ \hline
  $x$& \multicolumn{5}{c|}{Co$_2$[Mn$_{1-x}$Fe$_{x}$]Si} & \multicolumn{5}{c}{Co$_2$[Mn$_{1-x}$Fe$_{x}$]Sn}  \\
   &    Total  &   Co &   Mn&Fe&
 sp &  Total & Co &   Mn&Fe &  sp \\ \hline
  0.00  & 5.00&   1.96&    3.13  &          & -0.09&5.02&1.78&3.32&2.91& -0.08\\
  0.20  &  5.14&   2.13&     3.15 &    2.82 & -0.08&5.20&1.98&3.38&2.93& -0.07\\
  0.40  &  5.21&   2.25&     3.18&    2.79  & -0.07&5.37&2.19&3.44&2.93& -0.05\\
  0.60  &  5.25&   2.36&     3.20 &    2.78 & -0.05&5.47&2.36&3.45&2.93& -0.04\\
  0.80  &  5.29&   2.46&     3.23  &   2.78 & -0.04&5.52&2.49&3.53&2.94& -0.03\\
  1.00  &  5.28&   2.52&            &  2.79 & -0.03&5.52&2.61& &
  2.92&-0.01
\end{tabular}
\end{ruledtabular}
\end{table}

\begin{figure}
\includegraphics[scale=0.45]{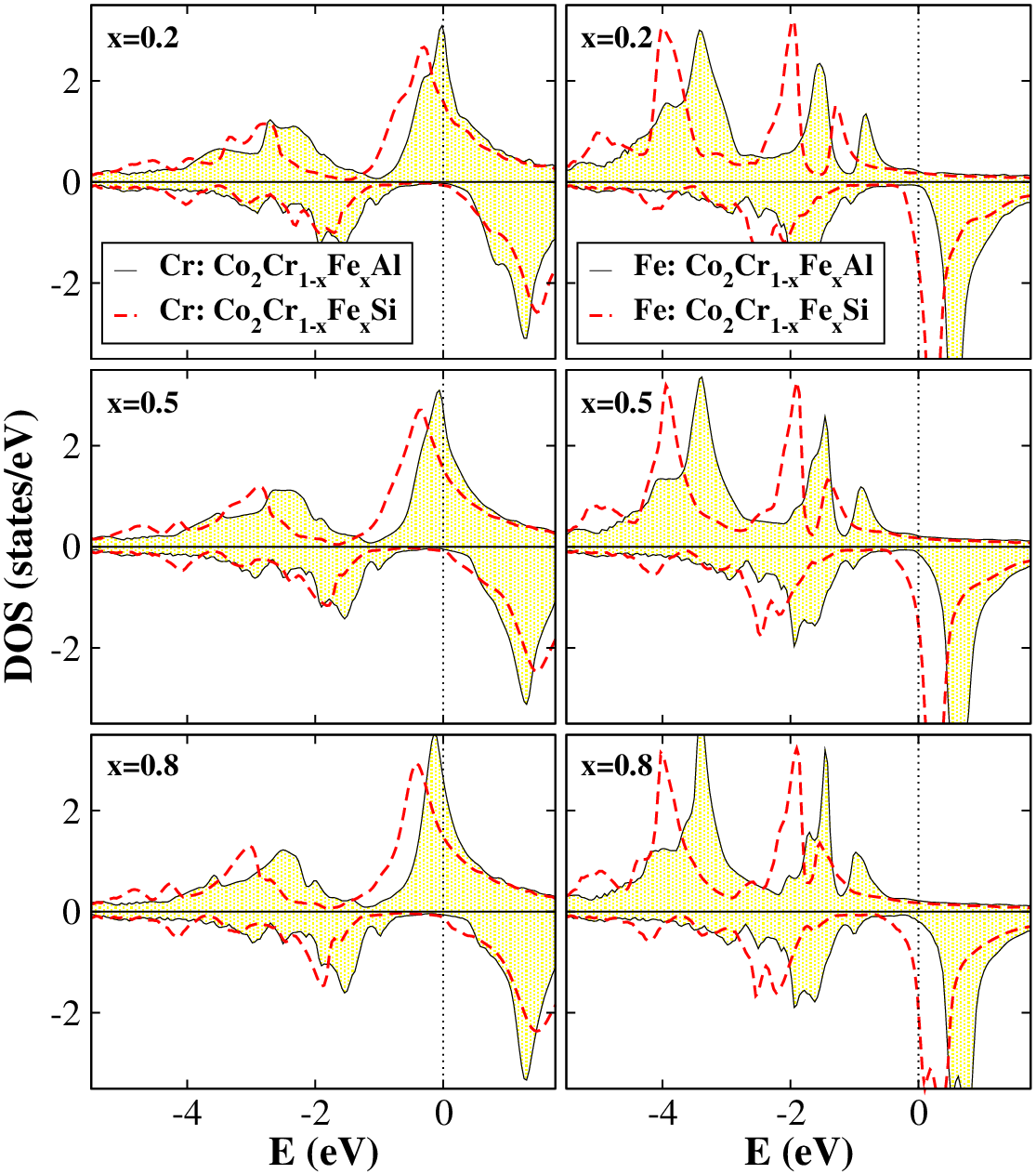}
\caption{(Color online) Cr- (left panel) and Fe-resolved (right
panel) DOS in the case of the Co$_2$[Cr$_{1-x}$Fe$_x$]Al and
Co$_2$[Cr$_{1-x}$Fe$_x$]Si compounds for three different values of
the concentration $x$. Details as in Fig. \ref{fig1}.
\label{fig9}}
\end{figure}

We will finish our investigation by examining the behavior of the
atoms in the different compounds. In Table \ref{table1} we have
included the atomic spin moments for the
Co$_2$[Cr$_{1-x}$Fe$_x$]Al (upper left panel) and
Co$_2$[Cr$_{1-x}$Fe$_x$]Si (upper right panel) and in Fig.
\ref{fig9} we have drawn the Cr-resolved DOS in the left panel and
the Fe-resolved one in the right panel for both Al- (solid black
lines with shaded area) and Si-based (red dashed lines) compounds
for three different values of the concentration $x$=0.2, 0.5 and
0.8. As we mentioned in the previous section, Co atoms increase
considerably their spin moments to account for the extra charge as
we dope with Fe. On the other hand Cr and Fe atoms have a pretty
constant spin moment through out all the concentrations range. Cr
has a spin moment of around 1.5 $\mu_B$ in the Al-based compound
and around 2.2 $\mu_B$ in the heavier Si-based compound. Fe on the
other hand in the Si compound cannot increase any more its spin
moment with respect to the Al compound and has a slight portion of
its minority states above the gap occupied and thus its spin
moment is slightly smaller for the heavier Si-compound than for
the Al-one and this is the main reason for the failure of the
compounds containing Fe and Si (or Ge or Sn) to retain the
half-metallicity.

The discussion in the previous paragraphs in this Section is
directly reflected on the atom-resolved DOS in Fig.~\ref{fig9}. Cr
in Co$_2$[Cr$_{1-x}$Fe$_x$]Al has an expected shape of DOS with
the Fermi level being pinned exactly at the maximum of the pick of
the majority states contrary to the case of the
Co$_2$[Mn$_{1-x}$Cr$_x$]Al compounds where it was below this
maximum and was shifted towards it as the Cr concentration
increased (see Ref,~\onlinecite{PRB-BR}). Substituting Si for Al
provides extra electrons and the majority states are pushed deeper
in energy being accompanied by a similar movement of the minority
states as in a rigid band model. But this shift is not strong
enough to lead to the loss of half-metallicity and the Fermi level
is located at the higher-energy-edge of the gap. These features
for the Si compound are similar to the ones of the perfect
Co$_2$CrSi studied recently by Chen et al.\cite{Podlucky} On the
other hand Fe majority states are deep in energy (Fe has two more
electrons than Cr) and for the Al-compound the Fermi level falls
at the higher-energy-edge of the minority gap which corresponds to
a region of very low majority DOS. When we pass from Al to Si
which has one valence electron more, the unoccupied states just
above the Fermi level cannot absorb the extra charge and it would
cost a lot in energy to occupy states far above the Fermi level
and thus the system prefers to occupy also minority states and to
loose its half-metallicity.

\section{Conclusions\label{sec4}}

We have complemented our study on the doping and disorder in
Co$_2$MnZ compounds presented in Refs. \onlinecite{APL} and
\onlinecite{PRB-BR} studying the quaternary
Co$_2$[Y$_{1-x}$Y$^*_x$]Z compounds with the lower-valent
transition metals Y,Y$^*$ being Cr, Mn or Fe and the sp atom Z
being one of Al, Ga, Si, Ge, Sn. Thus this study gave us a global
overview of the magnetic and electronic properties of these
compounds. The most important desirable feature of these compounds
is half-metallicity combined with high values of the majority
density of states at the Fermi level. The latter feature would
enable the good operation of the devices even when the
half-metallic character is lost due to defects, creation of
interfaces or temperature-driven phenomena. Our study shows that
the high values of majority density of states at the Fermi level
are ensured by the presence of Cr at high concentrations while the
only factor susceptible of destroying the half-metallicity is the
simultaneous presence of Fe and Si (or Ge or Sn) in the compound.
Moreover the presence of Cr or Mn and the exclusion of Fe is
responsible for larger minority band-gaps and thus for more stable
half-metallicity with respect to the phenomena which can induce
states within the gap.

To summarize, our results suggest that for realistic applications
the most appropriate compounds are the ones belonging to the
families Co$_2$[Mn$_{1-x}$Cr$_x$]Z with $x>0.5$ irrespectively of
the nature of $sp$ atoms since they combine high values of
majority DOS at the Fermi level, and half-metallicity with large
band-gaps. On the other hand the presence of Fe lowers
considerably the majority density of states at the Fermi level and
when combined with an element belonging to the Si-column, it even
can destroy half-metallicity.

\section*{Acknowledgements}
Authors  acknowledge  the computer support of the ``Leibniz
Institute for Solid State and Materials Research Dresden''.


\end{document}